\documentclass{elsart5p}
\usepackage{graphics}
\usepackage{graphicx}
\usepackage{epsfig}
\usepackage{amssymb}
\usepackage{amsmath}
\usepackage{color}

\newcommand{\dd}{\mbox{$\textrm{d}$}}

\begin{document}
\begin{frontmatter}

\title{Measurement of the absolute differential cross section of proton-proton elastic scattering at small angles}

\author[hepi,ikp]{D.~Mchedlishvili},
\author[hepi,ikp]{D.~Chiladze},
\author[jinr1,ikp]{S.~Dymov},
\author[hepi,ikp]{Z.~Bagdasarian},
\author[pnpi]{S.~Barsov},
\author[ikp]{R.~Gebel},
\author[cas,ikp]{B.~Gou},
\author[ikp]{M.~Hartmann},
\author[ikp]{A.~Kacharava},
\author[ikp]{I.~Keshelashvili},
\author[munster]{A.~Khoukaz},
\author[krakow]{P.~Kulessa},
\author[jinr1]{A.~Kulikov},
\author[ikp]{A.~Lehrach},
\author[hepi]{N.~Lomidze},
\author[ikp]{B.~Lorentz},
\author[ikp]{R.~Maier},
\author[hepi,jinr1]{G.~Macharashvili},
\author[jinr1,ikp]{S.~Merzliakov},
\author[ikp,pnpi]{S.~Mikirtychyants},
\author[hepi]{M.~Nioradze},
\author[ikp]{H.~Ohm},
\author[ikp]{D.~Prasuhn},
\author[ikp]{F.~Rathmann},
\author[ikp]{V.~Serdyuk},
\author[munster]{D.~Schroer},
\author[jinr1]{V.~Shmakova},
\author[ikp]{R.~Stassen},
\author[ikp]{H.J.~Stein},
\author[ikp]{H.~Stockhorst},
\author[wash]{I.I.~Strakovsky},
\author[ikp]{H.~Str\"oher},
\author[hepi]{M.~Tabidze},
\author[munster]{A.~T\"{a}schner},
\author[ikpros,msu]{S.~Trusov},
\author[jinr1]{D.~Tsirkov},
\author[jinr1]{Yu.~Uzikov},
\author[pnpi,bonn]{Yu.~Valdau},
\author[ucl]{C.~Wilkin\corauthref{cor1}}
\ead{c.wilkin@ucl.ac.uk} \corauth[cor1]{Corresponding author.},
\author[wash]{R.L.~Workman},
\author[zel]{P.~W\"{u}stner}
\address[hepi]{High Energy Physics Institute, Tbilisi State University, GE-0186 Tbilisi, Georgia}
\address[ikp]{Institut f\"ur Kernphysik, Forschungszentrum J\"ulich, D-52425
 J\"ulich, Germany}
\address[jinr1]{Laboratory of Nuclear Problems, JINR, RU-141980 Dubna, Russia}
\address[pnpi]{High Energy Physics Department, Petersburg Nuclear Physics
Institute, RU-188350 Gatchina, Russia}
\address[cas]{Institute of Modern Physics, Chinese Academy of Sciences, Lanzhou 730000, China}
\address[munster]{Institut f\"ur Kernphysik, Universit\"at M\"unster, D-48149
M\"unster, Germany}
\address[krakow]{H.~Niewodnicza\'{n}ski Institute of Nuclear Physics PAN, PL-31342
Krak\'{o}w, Poland}
\address[wash]{Data Analysis Center at the Institute for Nuclear Studies,
Department of Physics,\\
The George Washington University, Washington,D.C. 20052, USA}
\address[ikpros]{Institut f\"ur Kern- und Hadronenphysik,
Forschungszentrum Rossendorf, D-01314 Dresden, Germany}
\address[msu]{Department of Physics, M.~V.~Lomonosov Moscow State University,
RU-119991 Moscow, Russia}
\address[bonn]{Helmholtz-Institut f\"ur Strahlen- und Kernphysik, Universit\"at
  Bonn, D-53115 Bonn, Germany}
\address[ucl]{Physics and Astronomy Department, UCL, London WC1E 6BT, UK}
\address[zel]{Zentralinstitut f\"{u}r Engineering, Elektronik und Analytik, Forschungszentrum J\"{u}lich, D-52425 J\"{u}lich, Germany}

\date{\today}

\begin{abstract}
The differential cross section for proton-proton elastic scattering has been
measured at a beam energy of 1.0~GeV and in 200~MeV steps from 1.6 to 2.8~GeV
for centre-of-mass angles in the range from $12^{\circ}$--$16^{\circ}$ to
$25^{\circ}$--$30^{\circ}$, depending on the energy. Absolute normalisations
of typically 3\% were achieved by studying the energy losses of the
circulating beam of the COSY storage ring as it passed repeatedly through the
windowless hydrogen target of the ANKE magnetic spectrometer. It is shown
that the data have a significant impact upon a partial wave analysis. After
extrapolating the differential cross sections to the forward direction, the
results are broadly compatible with the predictions of forward dispersion
relations.
\end{abstract}

\end{frontmatter}

For beam energies above about 1~GeV there are relatively few measurements of
proton-proton elastic scattering at centre-of-mass (c.m.) angles $\theta$
from $10^{\circ}$ to $30^{\circ}$, i.e., between the region of major Coulomb
effects and the larger angles where the EDDA collaboration has contributed so
extensively~\cite{ALB1997,ALB2004,ALT2005}. This lack of information
inevitably leads to ambiguities in any $pp$ partial wave analysis (PWA) at
high energies~\cite{ARN2000}. The ANKE collaboration has recently published
proton analysing powers in this angular domain at 796~MeV and five other beam
energies between 1.6 and 2.4~GeV using a polarised proton beam~\cite{BAG2014}
and these led to a revision of the SAID PWA~\cite{ARN2000} in order to
accommodate the data. The major uncertainty in such a measurement is the
precision to which the beam polarisation can be determined, beam-target
luminosity and equipment acceptance playing only secondary roles. This is far
from being the case for the differential cross section where, in order to
provide accurate absolute values, both the luminosity and acceptance must be
mastered with high precision~\cite{CHI2011}. The difficulties encountered in
earlier experiments in making absolute measurements were discussed most
clearly in Ref.~\cite{SIM1993}, whose normalisation was used as the standard
for the EDDA work~\cite{ALB1997,ALB2004}.

As was the case for the analysing power~\cite{BAG2014}, the present studies
of the differential cross section were carried out using the ANKE magnetic
spectrometer~\cite{BAR2001} sited inside the storage ring of the COoler
SYnchrotron (COSY)~\cite{MAI1997} of the Forschungs\-zentrum J\"ulich. The
only detector used in the analysis was the forward one (FD), which measured
fast protons from elastic $pp$ scattering over a range of up to
$12^{\circ}-30^{\circ}$ in c.m.\ polar angles and $\pm 30^{\circ}$ in
azimuth. The FD comprises a set of multiwire proportional and drift chambers
and a two-plane scintillation hodoscope, the counters of which were used to
measure the energy losses required for particle
identification~\cite{DYM2004}.

The biggest challenge that has to be faced when measuring the absolute value
of a cross section in a storage ring experiment is to establish the
beam-target luminosity at the few percent level even though the overlap of
the beam and target cannot be deduced with such precision from macroscopic
measurements. It has been shown that this can be achieved by studying the
energy loss through electromagnetic processes as the coasting uncooled beam
passes repeatedly through the target chamber. There is a resulting change in
the frequency of the machine that can be determined with high accuracy by
studying the Schottky power spectrum of the beam~\cite{STE2008}. The amount
of electromagnetic interaction is, of course, proportional to that of the
strong proton-proton scattering, whose measurement was the goal of the
experiment

\begin{figure}[htb]
\centering
\includegraphics[width=1.0\columnwidth]{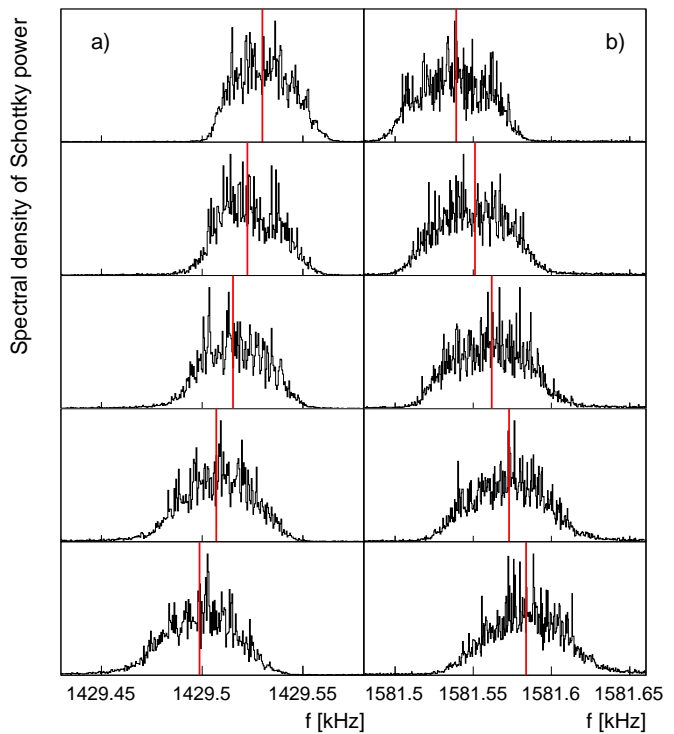}\hspace{5mm}
\caption{ Schottky power spectra obtained during one 300~s cycle and scaled
to harmonic number 1 for (a) 1.0 and (b) 2.0~GeV beam energies. Although
the data were recorded every 10 seconds, for
ease of presentation, only the results from every 60~s are shown, starting
from top to bottom. The mean frequencies are indicated by the vertical (red)
lines.} \label{fig:freqshift}
\end{figure}

The statistical distribution of particles in the beam is at the origin of the
Schottky noise. This gives rise to current fluctuations that induce a voltage
signal at a beam pick-up. The Fourier analysis of the voltage signal, i.e.,
of the random current fluctuations, by a spectrum analyser delivers power
distributions around the harmonics of the revolution frequency. Over a 300~s
cycle, the Schottky signals were recorded every 10~s with a 189~ms sweep
time, thus giving effectively instantaneous spectra. The frequencies were
measured with the existing Schottky pick-up of the stochastic cooling system
operated on harmonic number 1000~\cite{PRA2000} and a more precise analyser
than previously used~\cite{STE2008}. Some examples of these measurements
scaled to harmonic number 1 are shown in Fig.~\ref{fig:freqshift} for
circulating proton beam energies of 1.0 and 2.0~GeV. After subtracting the
background noise, the mean frequency $f$ of the beam at each instant of time
was evaluated from the centroid of the distribution. Such values, which are
indicated by the vertical lines, allow $f$ to be determined as a function of
time $t$ over the 300~s cycle.

The important point to notice in Fig.~\ref{fig:freqshift} is that the
direction of the frequency change is different at low and high energies; the
energy of this transition from one regime to the other depends upon the
lattice settings of the accelerator~\cite{STE2008}. Since the luminosity is
proportional to $\dd f/\dd t$, it means that there is a range of beam
energies where the fractional errors are so large as to make the Schottky
method of little practical use. This explains the gap in our data from
1.0~GeV to 1.6~GeV.

It was shown in Ref.~\cite{STE2008} that the effective number of target
particles is given by
\begin{equation}
 n_{T} = \left( \frac{1+\gamma}{\gamma}  \right)
         \frac{1}{\eta}
         \frac{1}{(\dd E/\dd x)m}
         \frac{T_p}{f^{2}}
         \frac{\dd f}{\dd t},
 \label{eq:nt}
\end{equation}
where $m$ is the proton mass. The cluster-jet target~\cite{KHO1999} used in
this experiment was very thin and, as a result, the energy changes over a
300~s cycle were extremely small ($\Delta E/E \approx 2-4\times 10^{-4}$).
Under such conditions one can take $f$ and $T_p$ to be the initial values of
the frequency and kinetic energy of the beam and $\gamma$ as the
corresponding Lorentz factor. The stopping power $\dd E/\dd x$ at a given
energy is to be found in the NIST-PML database~\cite{NIST}.

It is the remaining quantity in Eq.~(\ref{eq:nt}), the so-called
frequency-slip parameter $\eta$, that under COSY conditions changes sign at
$T_p\approx 1.3$~GeV. Although this parameter can be estimated
semi-quantitatively by a computer simulation of the acceleration process,
greater precision is achieved by a direct measurement, where the change in
the revolution frequency induced by adjusting the strength in the bending
magnets by few parts per thousand is studied~\cite{STE2008}. This was
investigated in separate runs at each of the beam energies with the target
switched off~\cite{CHI2011}.

A small frequency shift is also produced by the interaction of the beam with
the residual gas in the COSY ring and this was measured using dedicated
cycles, where the ANKE cluster target was switched on but the beam was
steered away from it. This precaution was necessary because the target
produces additional background in the vicinity of the ANKE target
chamber~\cite{STE2008}.

The measurement of the beam intensity ($n_B$) is a routine procedure for any
accelerator and is performed at COSY using the high precision Beam Current
Transformer device. These measurements were carried out every second over the
300~s cycle and then averaged. The final results are accurate to better than
$10^{-3}$~\cite{STE2008}. The luminosity in the experiment is then the
product of beam and target factors, $L=n_Bn_T$.

The Forward Detector was the subject of a very detailed study~\cite{DYM2015}
and only some of the salient points are mentioned here. The setup parameters
were adjusted in a geometry tuning procedure, with the use of the exclusive
$pp \to pp$, $pp\to pn\pi^+$, $pp\to pp\pi^0$, and $pp\to d\pi^+$ reactions.
In the last case, both the $d$ and $\pi^+$ were detected in the FD and this
gave valuable information on the systematics of the transverse momentum
reconstruction. These showed that any systematic shift in the determination
of the c.m.\ angle in $pp$ elastic scattering was less than $0.15^{\circ}$,
which would correspond to a $0.5\%$ change in the differential cross section.

The trigger was initiated by a signal from either of the two hodoscope walls,
placed one behind the other. This, together with a high scintillation counter
efficiency, reduced the trigger inefficiency to the $10^{-4}$ level. An MWPC
efficiency of over 97\% and a redundancy of the track hits led to an overall
tracking efficiency of $99.5\%$. In order to study the systematic effects,
different acceptance cuts were applied in the cross section evaluation as
well as different sets of wire planes used for track reconstruction. The
resulting values of the cross sections showed a $0.95\%$ (RMS) variation.

\begin{table}[h!]
\caption{\label{tab1}Percentage contributions to the total systematic
uncertainty at different proton beam energies $T_p$. E$_1$ reflects the
statistical and systematic effects in the determination of the Schottky
$\eta$ parameter. E$_2$ arises from the rest gas effect (including direct
measurement errors as well as possible instabilities). E$_3$ is a measure of
the density instability through the 300~s cycle. E$_4$ corresponds to the
accuracy of the stopping powers given in the NIST database~\cite{NIST}. E$_5$
is an estimate of the precision of the analysis of data taken with the FD.
These contributions have been added in quadrature to give the total
percentage uncertainty in the last column.} \vspace{2mm} \centering
\begin{tabular}{|c|c|c|c|c|c|c|}
\hline
$T_p$ &E$_1$&E$_2$&E$_3$&E$_4$&E$_5$&Total             \\
GeV   & \%  &  \% & \%  & \%  & \%  & \%                \\
\hline
1.0   &1.6&0.7&0.7&1.5&1.5&2.8\\
1.6   &1.2& 1.9 & 1.4 & 1.5 & 1.5 & 3.4 \\
\phantom{,}1.8\phantom{,}&\phantom{,}1.3\phantom{,}&\phantom{,}1.6\phantom{,}&\phantom{,}1.6\phantom{,}&\phantom{,}1.5\phantom{,}&\phantom{,}1.5\phantom{,}&\phantom{,}3.4\phantom{,}\\
2.0   &0.8 & 1.9 & 1.8 & 1.5 & 1.5 & 3.5 \\
2.2   &0.3 & 1.0 & 1.0 & 1.5 & 1.5 & 2.6 \\
2.4   &0.4 & 1.5 & 1.6 & 1.5 & 1.5 & 3.1 \\
2.6   &0.4 & 1.5 & 1.5 & 1.5 & 1.5 & 3.0 \\
2.8   &0.9 & 1.2 & 0.5 & 1.5 & 1.5 & 2.6 \\
\hline
\end{tabular}
\end{table}

The $pp$ elastic scattering reaction produced a prominent peak in the
missing-mass spectrum, with a background of only $1-2\%$, and an estimated
uncertainty of this level of $0.5\%$. A small contribution from the $pp\to
d\pi^+$ reaction to the peak region at 1~GeV was subtracted on the basis of
the energy-loss information. A kinematical fitting procedure was applied to
events in the $pp$ elastic scattering peak; this produced an accuracy
$\sigma(\theta_{\rm{cm}})= 0.1^{\circ}$ and further reduced the systematic
uncertainty.

Table~\ref{tab1} lists five identified contributions to the overall
systematic uncertainty of the $pp$ elastic scattering data at the eight
energies studied. No single contribution is dominant, which means that it
would be hard to reduce the systematic error to much below the 2.5--3.5\%
total uncertainty quoted in the table.

\begin{figure}[htb]
\centering
\includegraphics[width=1.0\columnwidth]{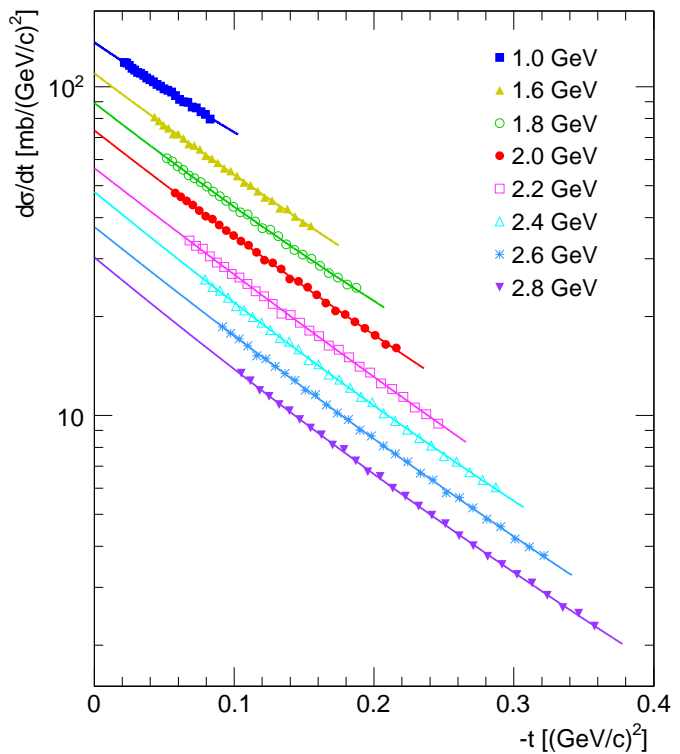}
\caption{Combined ANKE data set of differential cross sections with respect
to the four-momentum transfer $t$ compared to fits made on the basis of
Eq.~(\ref{pars}). Systematic errors are not shown. The correct values are
shown at 1.0~GeV but, for clarity of presentation, the other data are scaled
down sequentially in energy by factors of 1.2. The true numerical values of
the parameters are given in Table~\ref{tab2}. } \label{fig:dsdt}
\end{figure}

The variation of the ANKE data over the ranges in angle and energy studied
can be seen most clearly in the differential cross section with respect to
the four-momentum transfer $t$ and these results are shown in
Fig.~\ref{fig:dsdt}. Also shown are exponential fits to the measured data
made on the basis of
\begin{equation}
\label{pars}
\frac{\dd\sigma}{\dd t} = A\,\exp\left(-B|t|+C|t|^2\right),
\end{equation}
where the values of the resulting parameters are given in Table~\ref{tab2}.
Taking $C=0$ at 1~GeV would change the value of $A$ found in the fit by less
than 1\%, though this parameter becomes more important at higher energies
where the $t$ range is larger. This empirical representation of the measured
data may prove helpful when the results are used in the normalisation of
other experimental measurements.

\begin{table}[h!]
\caption{Parameters of the fits of Eq.~(\ref{pars}) to the differential cross
sections measured in this experiment. In addition to the statistical errors
shown, the second uncertainty in the value of $A$ in the second column
represents the combined systematic effects summarised in Table~\ref{tab1}.
The corrected values of the forward cross section, $A$(Corr.), were obtained
using the SAID fit discussed in the text, the associated error bars being
purely the systematic ones listed in Table~\ref{tab1}. These values, which
were not subjected to the SAID normalisation factors applied in
Fig.~\ref{fig:RON}, may be compared with those of $A$(GK), which were
determined using the Grein and Kroll forward amplitudes~\cite{GRE1982}.
\label{tab2}} \vspace{2mm} \centering
\begin{tabular}{|c|c|c|c|c|c|}
\hline
$T_p$ & $A$               & $B$              & $C$           &$A$(Corr.)  & $A$(GK) \\
GeV   & $\frac{\rm mb}{({\rm GeV}/c)^2}$  & (GeV$/c$)$^{\!-2}$ & (GeV$/c$)$^{\!-4}$&$\frac{\rm mb}{({\rm GeV}/c)^2}$&$\frac{\rm mb}{({\rm GeV}/c)^2}$\\
\hline
1.0   &$136.4\pm1.3\pm3.8$& $6.7\pm0.4$      &$4.0\pm3.8$    &$136.7\pm3.8$&$135.2$ \\
1.6   &$131.7\pm1.9\pm4.5$& $7.4\pm0.3$      &$2.7\pm1.7$    &$131.1\pm4.5$&$128.9$ \\
1.8   &$128.6\pm1.7\pm4.4$& $7.6\pm0.2$      &$3.4\pm1.0$    &$127.6\pm4.3$&$125.7$ \\
2.0   &$127.3\pm1.7\pm4.5$& $7.7\pm0.2$      &$2.5\pm0.8$    &$124.0\pm4.3$&$123.1$ \\
2.2   &$117.2\pm1.8\pm3.0$& $7.6\pm0.2$      &$1.4\pm0.7$    &$113.1\pm2.9$&$120.9$ \\
2.4   &$119.2\pm1.8\pm3.7$& $8.0\pm0.2$      &$2.7\pm0.5$    &$112.8\pm3.5$&$118.5$ \\
2.6   &$111.9\pm1.7\pm3.4$& $7.8\pm0.2$      &$2.0\pm0.4$    &$108.8\pm3.3$&$116.0$ \\
2.8   &$108.5\pm1.8\pm2.8$& $8.1\pm0.2$      &$2.4\pm0.4$    &$105.0\pm2.7$&$113.6$ \\
\hline
\end{tabular}
\end{table}

\begin{figure}[htb]
\centering
\includegraphics[width=1.0\columnwidth]{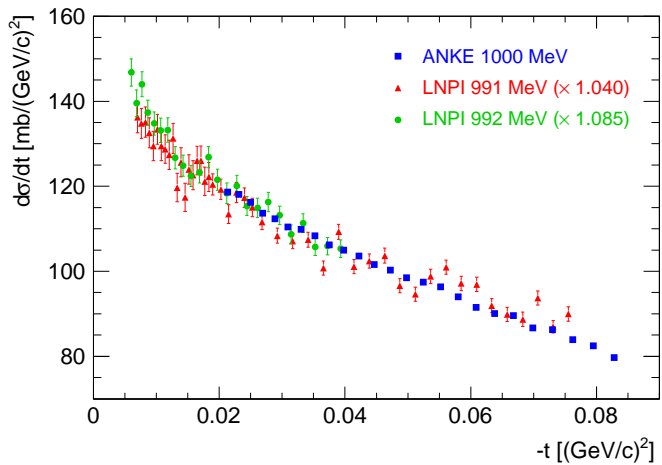}
\caption{Invariant differential cross section for $pp$ elastic scattering.
The ANKE data at 1000~MeV with statistical errors (blue squares) are compared
to the Gatchina hydrogen data at 992~MeV (green circles)~\cite{DOB1983}
scaled by a factor of 1.085 and methane results at 991~MeV (red
triangles)~\cite{DOB1988} scaled by a factor of 1.04. At very small values of
$|t|$ there is a rise caused mainly by Coulomb-nuclear interference. }
\label{fig:dsdt1000}
\end{figure}

There are very few data sets of absolute cross sections at small angles to
which the ANKE results can be compared. In the vicinity of 1~GeV there are
two measurements by the Gatchina group that were made with the IKAR recoil
detector. In the first of these at 992~MeV, IKAR was filled with
hydrogen~\cite{DOB1983}. In the second at 991~MeV a methane target was used,
though the prime purpose of this experiment was to show that such a target
gave consistent results and so could be used with a neutron
beam~\cite{SIL1989}. The ratio of the Gatchina hydrogen values~\cite{DOB1983}
to the fit of the ANKE 1000~MeV data over the common range of angles is
$0.920\pm 0.005$ and these Gatchina results have been scaled by a factor of
1.085 before being plotted in Fig.~\ref{fig:dsdt1000}. The scaling factor is
significant in view of the 2\% and 2.8\% absolute normalisations reported for
the Gatchina and ANKE experiments, respectively.

\begin{figure}[htb]
\centering
\includegraphics[width=1.0\columnwidth]{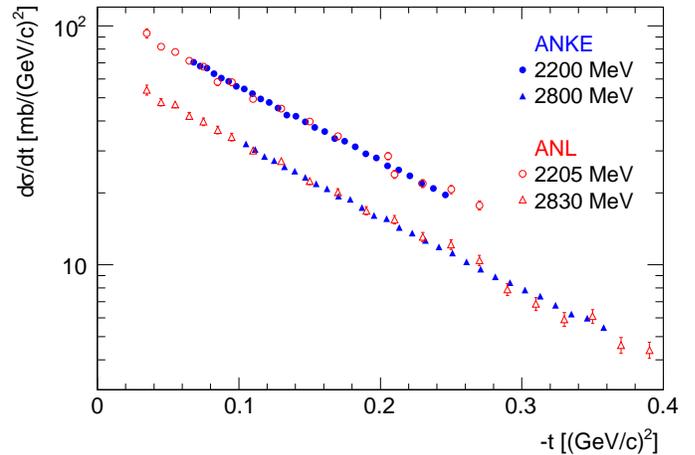}
\caption{The ANKE $pp$ differential cross section data at 2.2~GeV (closed
blue circles) and 2.8~GeV (closed blue triangles) compared to the ANL
results~\cite{AMB1974} at 2.2~GeV (open red circles) and 2.83~GeV (open red
triangles). Systematic errors are not shown. For presentational purposes,
both higher energy data sets have been scaled downwards by a common factor of
1.5.} \label{fig:Ambats}
\end{figure}

Data are also available from the Argonne National Laboratory in our angular
range at 2.2 and 2.83~GeV~\cite{AMB1974} and these values are plotted
together with our measurements in Fig.~\ref{fig:Ambats}. The ANL data sets
agree with our 2.2 and 2.8~GeV results to within 1\%. However, the absolute
normalisation claimed for these data was 4\%~\cite{AMB1974} so that it is not
possible to draw completely firm conclusions from this comparison.

\begin{figure}[htb]
\centering
\includegraphics[width=1.0\columnwidth]{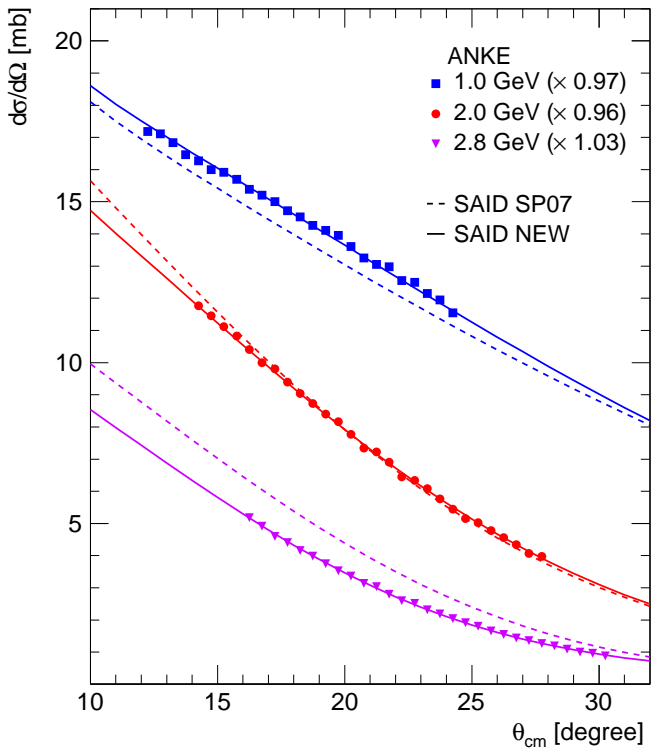}
\caption{Scaled ANKE proton-proton elastic differential cross sections at
1.0, 2.0, and 2.8~GeV with statistical errors compared to the SAID SP07
solution~\cite{ARN2000} and a modified (``new'') partial wave solution where
the ANKE data have been taken into account. For presentational reasons the
2.0 and 2.8~GeV data and curves have been reduced by factors of 0.5 and 0.25,
respectively. The best agreement with the new partial wave data was achieved
by scaling the ANKE data with factors 0.97, 0.96, and 1.03 at the three
energies. Such factors are within the uncertainties given in
Table~\ref{tab1}.} \label{fig:RON}
\end{figure}

The results reported in this letter could clearly have an impact on the
current partial wave solutions. This is demonstrated in Fig.~\ref{fig:RON},
where the ANKE cross sections at 1.0, 2.0, and 2.8~GeV are compared to both
the SAID SP07 solution~\cite{ARN2000} and a modified one that takes the
present data at all eight energies into account. Scaling factors in the
partial wave analysis, consistent with the overall uncertainties given in
Table~\ref{tab1}, have been included in the figure. The major changes
introduced by the new partial wave solution are in the $^{1\!}S_0$ and
$^{1\!}D_2$ waves at high energy. It should be noted that the modified
solution does not weaken the description of the ANKE proton analysing powers
presented in Ref.~\cite{BAG2014}.

The precise EDDA measurements were undertaken for c.m.\ angles of
$35^{\circ}$ and above whereas the ANKE data finish well below this and the
gap looks even bigger in terms of the momentum-transfer variable $t$.
Nevertheless, the modified SAID solution shown in Fig.~\ref{fig:RON} fits the
ANKE 1000~MeV cross section reduced by 3\% and describes also the EDDA data
at 1014.4~MeV~\cite{ALB2004}. Such a 3\% reduction in the ANKE normalisation
at this energy is consistent with the results of a combined fit of
Eq.~(\ref{pars}) to the EDDA and the Coulomb-corrected ANKE data.

\begin{figure}[htb]
\centering
\includegraphics[width=1.0\columnwidth]{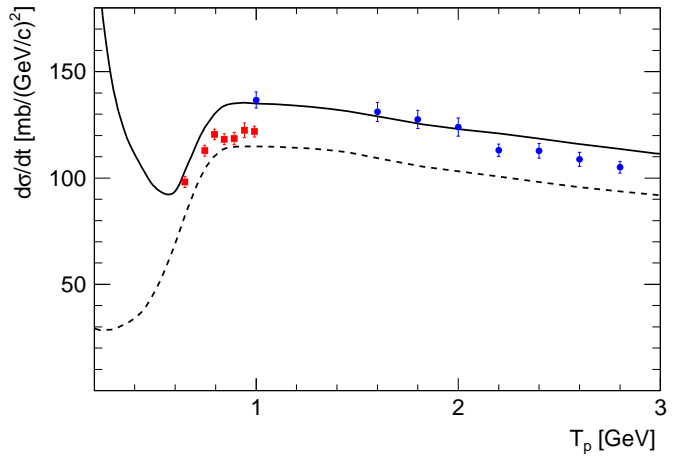}
\caption{The predictions of Grein and Kroll~\cite{GRE1982} for the values of
the forward $pp$ elastic differential cross section (solid line), the
corresponding lower limit provided by the spin-independent optical theorem of
Eq.~(\ref{opt}) being indicated by the broken line. The extrapolated ANKE
data, corresponding to the $A$(Corr.) parameter of Table~\ref{tab2}, are
shown with their quoted errors by the (blue) circles, whereas the (red)
squares are the published Gatchina values~\cite{DOB1983}.} \label{fig:GK}
\end{figure}

In the forward direction the number of proton-proton elastic scattering
amplitude reduces from five to three and the imaginary parts of these
amplitudes are determined completely by the spin-averaged and spin-dependent
total cross sections with the help of the generalised optical theorem. The
corresponding real parts have been estimated from forward dispersion
relations, where these total cross sections provide the necessary
input~\cite{GRE1982}. All the terms contribute positively to the value of $A$
and, using the optical theorem, the lower bound,
\begin{equation}
\label{opt}
A\ge (\sigma_{\rm tot})^2/16\pi,
\end{equation}
is obtained by taking the $pp$ spin-averaged total cross section $\sigma_{\rm
tot}$. This lower bound and the full Grein and Kroll estimates for
$A$~\cite{GRE1982} are both shown in Fig.~\ref{fig:GK} where, for
consistency, the same values of $\sigma_{\rm tot}$ were used in the two
calculations.

The 992~MeV Gatchina data of Fig.~\ref{fig:dsdt1000} show a significant rise
at small $|t|$ that is a reflection of Coulomb distortion of the strong
interaction cross section and this was taken into account through the
introduction of explicit corrections~\cite{DOB1983}. The corrected data were
then extrapolated to the forward direction ($t=0$), using a simple
exponential function, which would correspond to Eq.~(\ref{pars}) with $C=0$.
The resulting points at all the energies studied are generally about 10\%
below the Grein and Kroll predictions and would therefore correspond to
smaller real parts of the spin-dependent amplitudes.  The extrapolation does,
of course, depend upon the Coulomb-corrected data following the exponential
fit down to $t=0$.

Though the ANKE data do not probe such small $|t|$ values as
Gatchina~\cite{DOB1983}, and are therefore less sensitive to Coulomb
distortions, these effects cannot be neglected since they contribute between
about 1.5\% and 4.5\% at 1.0~GeV though less at higher energies. It is seen
in Fig.~\ref{fig:RON} that modified SAID solutions describe well the ANKE
measurements at three typical energies and the same is true also at the
energies not shown. After fitting the ANKE measurements, there is a facility
in the SAID program for switching off the Coulomb interaction without
adjusting the partial wave amplitudes~\cite{ARN2000} and this allows a robust
extrapolation of the Coulomb-free cross section to the forward direction. The
approach has the advantage that it includes some of the minor Coulomb effects
that are contained in the SAID program~\cite{ARN1983,LEC1980}. It takes into
account the phase variations present in the partial wave analysis and also
the deviations from exponential behaviour for very small momentum transfers,
$|t| \lesssim m_{\pi^0}^2 = 0.018$~(GeV/$c$)$^2$, that are linked to pion
exchange. The values for $A$(Corr.) at $t=0$ produced in this way are given
in Table~\ref{tab2} and shown in Fig.~\ref{fig:GK}. The error bars are purely
the systematic uncertainties listed in Table~\ref{tab1} and any errors in the
angular dependence of the SAID predictions are neglected.

The corrections obtained using the SAID program with and without the Coulomb
interaction at 1~GeV are a little larger than those found by the Gatchina
group using an explicit Coulomb formula~\cite{DOB1983}, in part due to the
different relative real parts of the $pp$ amplitude in the two calculations.

The agreement of the ANKE data with the theoretical curve in
Fig.~\ref{fig:GK} is encouraging and would be even slightly better if the
normalisation factors found in the fits to the cross sections in
Fig.~\ref{fig:RON} were implemented. Nevertheless, the extrapolated values
generally fall a little below the predictions at the higher energies.

In summary we have measured the differential cross sections for proton-proton
elastic scattering at eight energies between 1.0 and 2.8~GeV in a c.m.\
angular domain between about $12^{\circ}$--$16^{\circ}$ to
$25^{\circ}$--$30^{\circ}$, depending on the energy. Absolute normalisation
of typically 3\% were achieved by measuring the energy loss of the beam as it
traversed the target. After taking the Coulomb distortions into account, the
extrapolations to the forward direction, are broadly compatible with the
predictions of forward dispersion relations.

Although our results are completely consistent with ANL measurements at 2.2
and 2.83~GeV~\cite{AMB1974}, the published Gatchina values~\cite{DOB1983} are
lower than ours at 1~GeV by about 8\%, though this would be reduced to about
5\% if one accepts the renormalisation factor from the SAID fit shown in
Fig.~\ref{fig:RON}.

The new ANKE data have a significant influence on a partial wave analysis of
this reaction, changing in particular $^{1\!}S_0$ and $^{1\!}D_2$ waves at
high energies. This will be made clearer in an update to the SAID SP07
solution~\cite{ARN2000}. On a more practical level, the measurements will
also be a valuable tool in the normalisation of other experiments.

We acknowledge valuable discussions with Peter Kroll, who provided the
numerical values of the predictions shown in Ref.~\cite{GRE1982}. We also are
grateful to other members of the ANKE Collaboration for their help with this
experiment and to the COSY crew for providing such good working conditions.
This material is based upon work supported by the Forschungszentrum J\"ulich
(COSY-FEE), the Shota Rustaveli Science Foundation Grant (\#31/91), and the
U.S. Department of Energy, Office of Science, Office of Nuclear Physics,
under Award Number DE-FG02-99-ER41110.

\newpage

%
%

%

\end{document}